\documentclass[aps, prl, reprint, floatfix, superscriptaddress]{revtex4-1}
\newcommand{\ino}[0]{$\mathrm{In_2 O_3}$}
\usepackage[version=3]{mhchem}
\usepackage{siunitx}
\usepackage{hhline}
\usepackage{placeins}
\usepackage[caption=false]{subfig}

\usepackage{color,graphicx}
\newcolumntype{M}[1]{>{\centering\arraybackslash}m{#1}}

\begin{document}
\title{Structures and finite-temperature abundances of defects in
  In$_2$O$_3$-II from first-principles calculations} 
  \author{Jamie M. Wynn} 
\affiliation{Centre for Scientific Computing, Cavendish Laboratory, University of Cambridge}
\affiliation{Theory of Condensed Matter Group, Cavendish Laboratory, University of Cambridge}
  \author{Richard J. Needs} 
\affiliation{Theory of Condensed Matter Group, Cavendish Laboratory, University of Cambridge}
  \author{Andrew J. Morris}
\affiliation{Theory of Condensed Matter Group, Cavendish Laboratory, University of Cambridge}

\begin{abstract}
  We have studied intrinsic defect complexes in \ce{In2O3}
  using \emph{ab initio} random structure searching (AIRSS).  Our
  first-principles density-functional-theory calculations predict the
  thermodynamic stability of several novel defect structures.  We
  combine the static lattice energy and harmonic vibrational energy
  with the often-neglected configurational entropy to construct the free energy,
  which is minimised to predict defect abundances at finite
  temperatures.  We predict that some of our new defect structures -- 
  in particular our \{In,2\emph{V}$_\mathrm{O}$\} and \{2In,3\emph{V}$_\mathrm{O}$\} defects -- can exist in significant abundances at finite temperatures, and their
  densities of electronic states indicate that they could play an important role 
  in the unexpectedly high density of n-type charge carriers
  observed in \ino{}.
\end{abstract}

\maketitle


The thermodynamically most stable crystalline form of indium oxide
(\ino{}) is a transparent semiconducting oxide (TCO) that adopts the
bixbyite structure. TCOs are relatively uncommon materials, but they
are of significant technological interest. They combine the usually
mutually exclusive properties of optical transparency in the visible
range with high electrical conductivity.  It has been observed that
nominally undoped \ino{} exhibits intrinsic n-type conductivity, which
has been associated with the presence of both oxygen vacancy
defects~\cite{Curreli2005,Bierwagen2015} and hydrogen
impurities~\cite{Galazka2013}. The aim of this work is to study the
energetics and structures of low-energy defects in undoped \ino{}.

Undoped \ino{} exhibits n-type conductivity with carrier
concentrations of up to
$\SI{e19}{\centi\metre^{-3}}$~\cite{DeWit1977}, even when grown in
exceptionally pure environments~\cite{Bierwagen2015}. This phenomenon
is known as unintentional doping (UID).  The UID in $\mathrm{In_2
  O_3}$ was originally attributed to oxygen vacancies acting
effectively as donors~\cite{DeWit1977}; consistently with this, \ino{}
is a non-stoichiometric material with an excess of In relative to
O~\cite{DeWit1977}. However, studies have been inconsistent on whether
such vacancies are sufficiently shallow in energy to explain the
effect~\cite{Limpijumnong2009, Bierwagen2012}. Other candidates which
have been put forward to explain the unexpectedly high carrier density
of \ino{} include indium self-interstitials~\cite{Tang2010} and
hydrogen impurities~\cite{King2009}. Experimentally, the
carrier density of \ino{} depends strongly on the temperature at which
it is manufactured~\cite{Galazka2013}.  A great deal of research has
been focused on optoelectronic applications of TCOs created by doping
$\mathrm{In_2 O_3}$ with tin~\cite{Agoston2009}, which has been used
in energy-efficient windows~\cite{Hamberg1986} and as an anode in
OLEDs (organic light-emitting diodes)~\cite{ParkKwakKimEtAl2002}.
Undoped \ino{} also finds technological applications in gas
sensors~\cite{Xu2013}, and \ino{} nanotubes \cite{Shen2005} have
applications in arrays of nanotubes for biosensing \cite{Curreli2005}.

Density-functional theory (DFT) has previously been used to elucidate
the optical properties of \ino{} that give rise to its
transparency~\cite{Walsh2008} and to calculate formation energies of
intrinsic point
defects~\cite{AgostonAlbeNieminenEtAl2009,Agoston2009}. In this work
we go beyond previous studies by predicting the structures of defect
complexes from first principles, which has allowed us to identify many new
defects.  We have used the \textit{ab initio} random structure
searching method (AIRSS) \cite{Pickard2011} to discover low energy
defect structures.  We have also incorporated finite-temperature
effects in calculating their energetic stability.  Previously AIRSS has been used
to study phases of materials at high pressures~\cite{Pickard2006},
to determine structures of point defect complexes formed by
impurities in silicon~\cite{Morris2011,Morris2009,Morris2013}, to
investigate zirconolite as a material for nuclear waste
encapsulation~\cite{Mulroue2011,Mulroue2013}, and to map out the
sequence of phase transitions in the lithiation and sodiation of
phosphorus anodes in batteries~\cite{Mayo}.

Our calculations provide the static lattice energies, the
configurational entropy accounting properly for the symmetries of
defects, and the contribution of the phonons to the free energy. Each
of these contributions can have a large impact on the defect
abundances at ambient conditions. All DFT calculations reported in
this letter were performed using the \textsc{castep} plane-wave
code~\cite{Clark2005}; the methodological details are provided in the
Supplemental Material.

\label{sec:compmethods}
 
The bixbyite structure~\cite{Karazhanov2007} of $Ia\bar{3}$ symmetry,
also known as \ce{In2O3}-II, is thermodynamically stable at ambient
conditions. The structure contains two inequivalent types of In atom
on the $8b$ and $24d$ Wyckoff sites. All oxygen atoms are equivalent
and are on the $48e$ site.

In AIRSS a large number of random structures are generated using DFT
methods and relaxed to the nearest local minimum of the energy
landscape~\cite{Pickard2011}. The structures that are particularly low
in energy can then be singled out for further examination.  The
structure searching computations were performed using the 40-atom
primitive cell of \ino{}-II, and more accurate calculations on the
most promising defects were performed in a 160-atom supercell.

The initial structures for the AIRSS searches were created by selecting
a sphere of radius \SI{3}{\angstrom} centred on a random position
within the \ino{} unit cell, and then removing atoms within the sphere
at random and inserting In or O atoms. The atomic coordinates of all
atoms within the sphere were randomised, making sure that they were
not unphysically close to one another.

We adopt throughout a notation in which defects are indicated by
listing their contents within braces; we also use $V_\mathrm{x}$ to
refer to a vacancy of element x so that, for example, \{In,
$V_\mathrm{O}$\} constitutes an indium substitutional. Metastable
defects are indicated using asterisks; for example, the lowest-energy
oxygen vacancy would be denoted using \{$V_\mathrm{O}$\}, the
second-lowest as \{$V_\mathrm{O}$\}*, and the third-lowest as
\{$V_\mathrm{O}$\}**.

\label{sec:method}

Because \ino{} empirically exhibits a deficit of oxygen relative to
its ideal bulk stoichiometry, we searched for defects formed by a mixture
of oxygen vacancies and indium interstitials. These searches generated
a total of 632 defect structures (although many of them are
duplicates).  To our knowledge, none of the `nontrivial' defects found
here -- \emph{i.e.}, anything but single-atom interstitials,
vacancies, and substitutionals -- has been reported previously in the
literature.

\label{sec:futs}

Defect abundances at finite temperatures are calculated by
constructing the free energy and minimising it with respect to the
abundances. Our expression for the free energy takes into account
contributions from the static-lattice formation energies, phonons, and
the configurational entropy associated with the defects:
\begin{align}
\begin{split}
  F & = \sum\limits_{i} n_i E_i + \sum\limits_{i} n_i F_i^{\mathrm{vib}} \\
  & - k_B T \ln\left(\frac{N!\prod_i w_i^{n_i}}{\left(N - \sum_i
        n_i\right)!\prod_i n_i!}\right),
\label{eq:futs}
\end{split}
\end{align}
\noindent where the $n_i$ are the number of defects of type $i$, $N$
is the total number of lattice sites, $w_i$ is the configurational
degeneracy per lattice site, $F_i^\mathrm{vib}$ is the
temperature-dependent vibrational free energy, and $E_i$ is the
formation energy of the $i$-th
defect.

The vibrational free energy is given within the harmonic approximation
by~\cite{Al-MushadaniNeeds2003}
  ${F_i^\mathrm{vib} = k_B \int g_i(\omega)\ln(2\sinh(\beta\hbar\omega))\mathrm{d}\omega}$,
where $g_i(\omega)$ is the phonon density of states of the $i$-th
defect.  We calculate the configurational degeneracies $w_i$ of the
defects by applying each symmetry operator of the space group of bulk
\ino{} and counting the number of differently oriented versions of the
same defect that are thereby generated. (In the calculation of the
$w_i$, we count defects which differ only by a translation as the
same, since the entropy associated with placing a defect on different
lattice sites is already included in Eq.~\ref{eq:futs}.) The logarithm
of the degeneracy then gives the orientational configurational entropy
per lattice site (divided by $k_B$), as described in
Ref.~\citealp{Morris2008}. Along with a contribution from the
combinatorics of assigning the defects to particular lattice sites,
this forms the third term of Eq.~\ref{eq:futs}.
The defect abundances are computed by minimising Eq.~\ref{eq:futs}
with respect to the $n_i$ at some particular temperature, $T$.

\label{sec:oxvac}

\begin{figure}
    \centering
    \includegraphics[width=1.0\columnwidth]{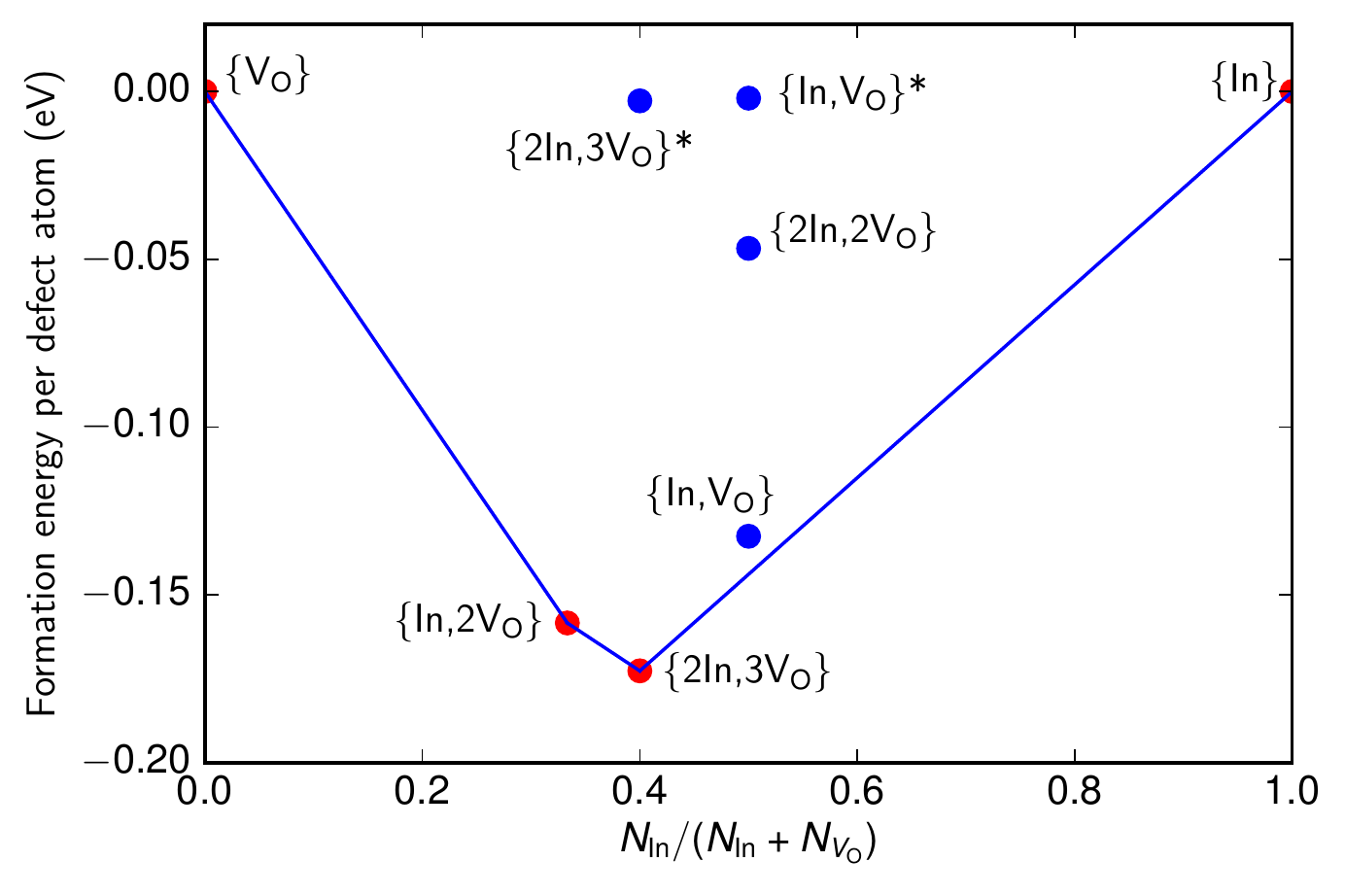}
    \caption{Convex hull for defect complexes in \ino{} consisting of
      oxygen vacancies and indium interstitials. The convex hull is
      shown in blue. Defects lying on the hull are red, and those
      above the hull are blue.  }
    \label{fig:oxvachull}
\end{figure}

\begin{figure*}
\centering
\hfill
    \subfloat[]{
\includegraphics[width=0.30\linewidth]{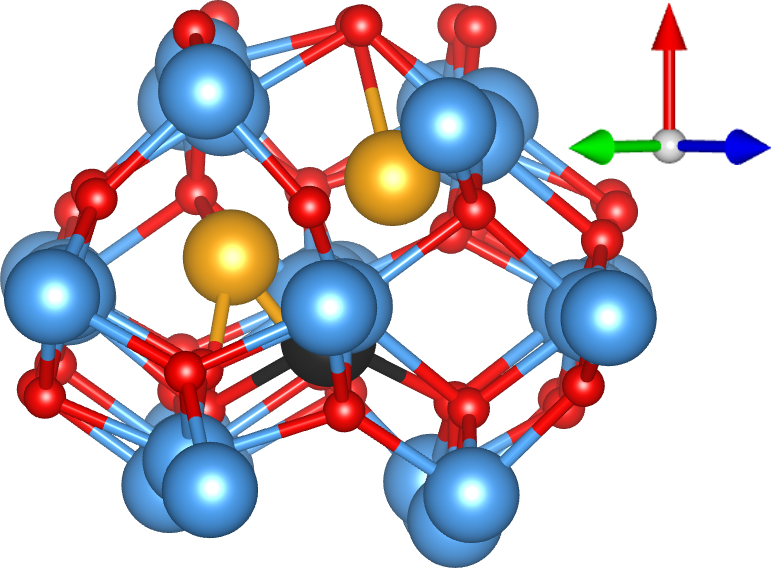}
}
\hfill
    \subfloat[]{
\includegraphics[width=0.30\linewidth]{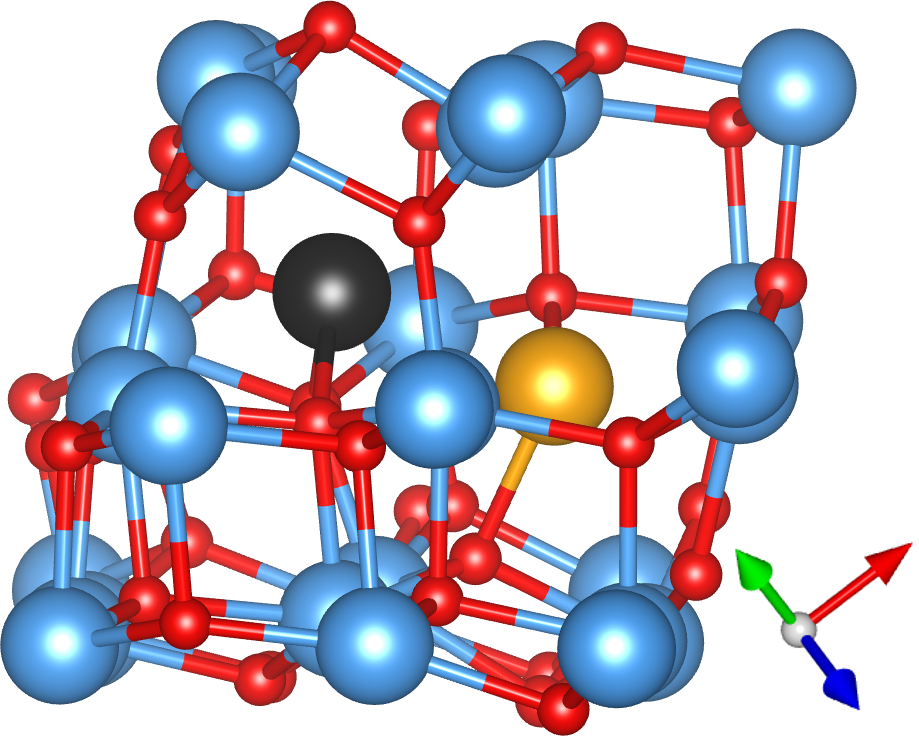}
}
\hfill
    \subfloat[]{
\includegraphics[width=0.33\linewidth]{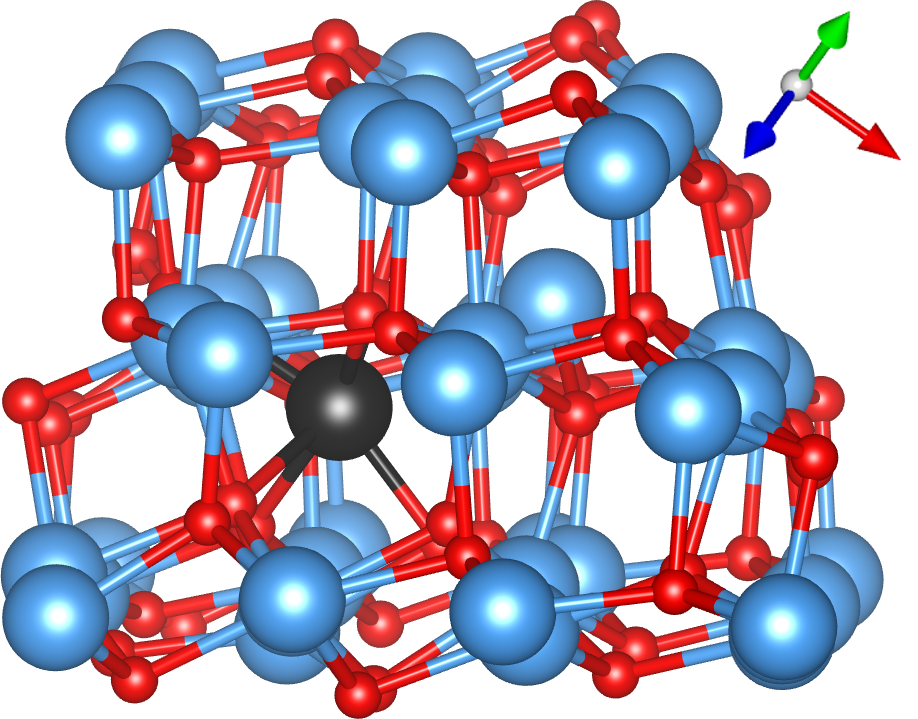}
}
\hfill\null
    \caption{Left to right, the \{2In,3\emph{V}$_\mathrm{O}$\}, \{In,2\emph{V}$_\mathrm{O}$\} and \{In,\emph{V}$_\mathrm{O}$\} defects.
      For visual clarity we show only the structure in the vicinity of
      the defect, rather than the entire unit cell. Indium atoms are
      shown in blue and oxygen atoms in red. (Black and orange indium
      atoms are referred to specifically in the text.) Lattice vectors
      of the conventional cubic unit cell are shown next to each
      defect, although they are not labelled because under the
      $Ia\bar{3}$ space group the three lattice vectors are
      symmetrically equivalent.}
    \label{fig:defectstructures}
\end{figure*}

The formation energy of a defect is defined as~\cite{Freysoldt2014a}
\begin{equation}
E_F = E_D - \sum_\alpha N_\alpha \mu_\alpha - E_B.
\label{eq:ef}
\end{equation}
In the above, $N_\alpha$ is the number of defect atoms of species
$\alpha$; the $\mu_\alpha$ are the associated chemical potentials;
$E_D$ is the total energy of the defective unit cell; and $E_B$
is the total energy of the bulk unit cell. 

To visualise the energetics of the binary defects generated by the
structure searching, we use the Maxwell construction, which has
previously been applied to defects in Ref.~\citealp{Morris2011}.  The
Maxwell construction is obtained by plotting the formation energy per
atom of each defect against its composition.  At $T=0$, only those
defects lying on the convex hull of the resulting scatter plot are
thermodynamically stable.

The Maxwell construction for the most energetically favorable defects
after re-optimisation in the larger 160-atom unit cell is shown in
Fig.~\ref{fig:oxvachull}.
Note that we treat vacancies just as we would atoms, so that in
Fig.~\ref{fig:oxvachull}, the $x$-axis runs from the case where there
are oxygen vacancies but no indium interstitials to the case where
there are indium interstitials but no oxygen
vacancies. Fig.~\ref{fig:oxvachull} shows how the two types of defect
bind as a function of their concentrations relative to each other.
The chemical potentials for $V_\mathrm{O}$ and In were calculated
using the oxygen vacancy and indium interstitial defects, with
$\mu_{V_\mathrm{O}} = E(\{V_\mathrm{O}\}) - E_B$.  (We note that the
convex hull ceases to be meaningful at $T>0$, since Eq.~\ref{eq:futs}
shows that the entropy is a nonlinear function of the defect
concentrations.)

We now survey the structures of the defects whose energies are
reported in Fig.~\ref{fig:oxvachull}. The lowest-energy indium
interstitial was found to be the $c$-site defect previously reported
in Ref.~\citealp{Agoston2009}. Since all oxygen atoms in \ino{} occupy
the same Wyckoff site, there is only one way to remove an oxygen atom,
leading to a single possible O vacancy defect (barring significant
disruption to the lattice in its vicinity, which we find to be
energetically unfavorable).

The defect with the lowest formation energy was found to be
\{2In,3\emph{V}$_\mathrm{O}$\}, which is shown in
Fig.~\ref{fig:defectstructures}(a). In this defect, two In atoms
(orange) sit either side of an In site as in a split
interstitial. This is accommodated by the removal of 3 O atoms and the
significant displacement of another In atom (black) away from the
defect.  The \{In,2\emph{V}$_\mathrm{O}$\} defect shown in
Fig.~\ref{fig:defectstructures}(b) is also on the convex hull and
consists of an interstitial indium atom (black) with partially ionic
bonds to two nearby oxygen atoms, which significantly displace a
nearby indium atom (orange, to its lower-right). The
\{In,\emph{V}$_\mathrm{O}$\} defect, consisting of an In
substitutional as shown in Fig.~\ref{fig:defectstructures}(c), is
close enough to the hull to become potentially relevant at finite
temperatures, given the intrinsic errors in DFT calculations. In this
defect, an In atom (black) substitutes for an O atom, which
significantly displaces the closest In atom, which is visible to its
top-right in Fig.~\ref{fig:defectstructures}(c).  To our knowledge,
these defect structures have not been reported previously in the
literature except for the indium interstitial and oxygen vacancy at
the endpoints of the convex hull from which the chemical potentials
were taken.  The metastable \{2In,2\emph{V}$_\mathrm{O}$\},
\{In,\emph{V}$_\mathrm{O}$\}*, and \{2In,3\emph{V}$_\mathrm{O}$\}*
defects shown in Fig.~\ref{fig:oxvachull} are described in the
Supplemental Material.

\begin{figure}
    \centering
    \includegraphics[width=1.0\columnwidth]{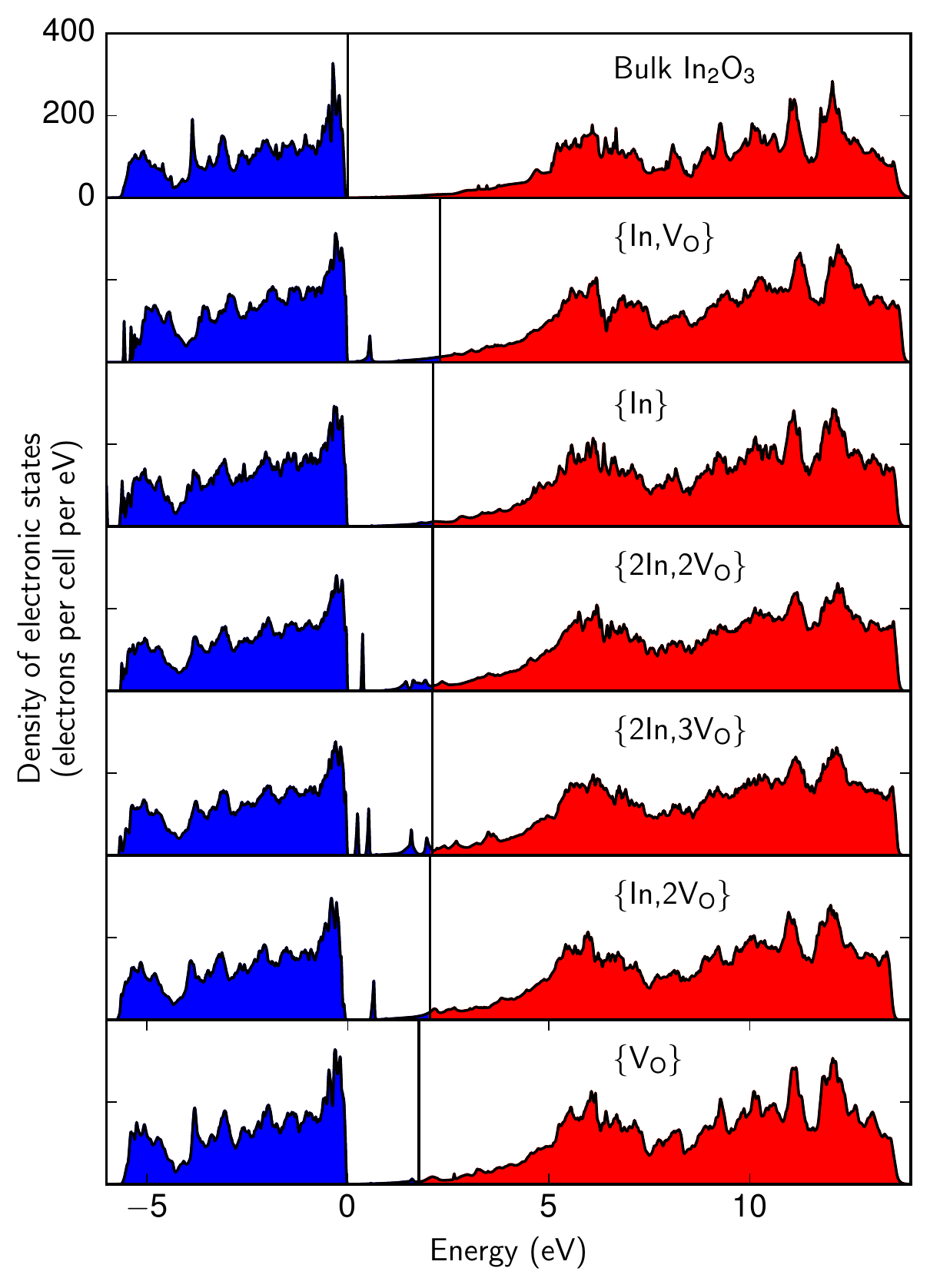}
    \caption{Electronic density of states (DOS) of the low-lying \ino{}
      defects shown in Fig.~\ref{fig:oxvachull}, together with the
      bulk density of states for comparison. Above the VBM, each DOS
      has been multiplied by a factor of 2 for visual clarity. The
      Fermi level is shown in each case by a vertical black line. All
      DOS have been shifted in order to align their VBMs. Occupied
      states are shown in blue and unoccupied states are shown in
      red. Tick marks on each DOS match those shown at the top.}
\label{fig:oxvacdos}
\end{figure}

Having identified the structures of defect complexes formed from
oxygen vacancies and indium interstitials, their capacity to act as
shallow donors was studied by calculating their densities of
electronic states (DOS); if any of the defects identified create
occupied states close to the conduction band, they are potentially
implicated in the intrinsic n-type doping that is observed in \ino{}.
We investigated the DOS of each defect
lying close to the convex hull, and obtained the results shown in
Fig.~\ref{fig:oxvacdos}, given alongside the bulk DOS for reference.
This was done for each defect lying close to the convex hull. The
densities of states were calculated using \textsc{castep} and the
adaptive broadening functionality~\cite{Yates2007} of the
OptaDOS~\cite{Morris2014} code.

As is typical, the PBE functional significantly underestimates the
band gap of \ino{}; our calculations produce a gap of
\SI{0.8}{\electronvolt}, compared to an experimental value of around
\SI{3}{\electronvolt}. It is therefore possible, in principle, that
the presence of occupied defect levels near the bottom of the valence
band is simply the result of the inaccurately small gap predicted by
the PBE functional. We tested this by recalculating the 
DOS for the oxygen vacancy using the
HSE06~\cite{KrukauVydrovIzmaylovEtAl2006} screened exchange functional
which gives larger and more accurate band gaps than PBE.
By comparing the densities of states with the PBE
and HSE06 functionals, we find that using an accurate band gap shifts
the defect levels shift upwards with the conduction band rather than
being fixed with respect to the valence band. This dispels concerns
that the shallowness of the defect levels relative to the valence band
may be an artifact of the unphysically small PBE band gap. (The two
densities of states are shown in the Supplemental Material.)

\begin{figure*}
    \centering
    \includegraphics[width=0.9\textwidth]{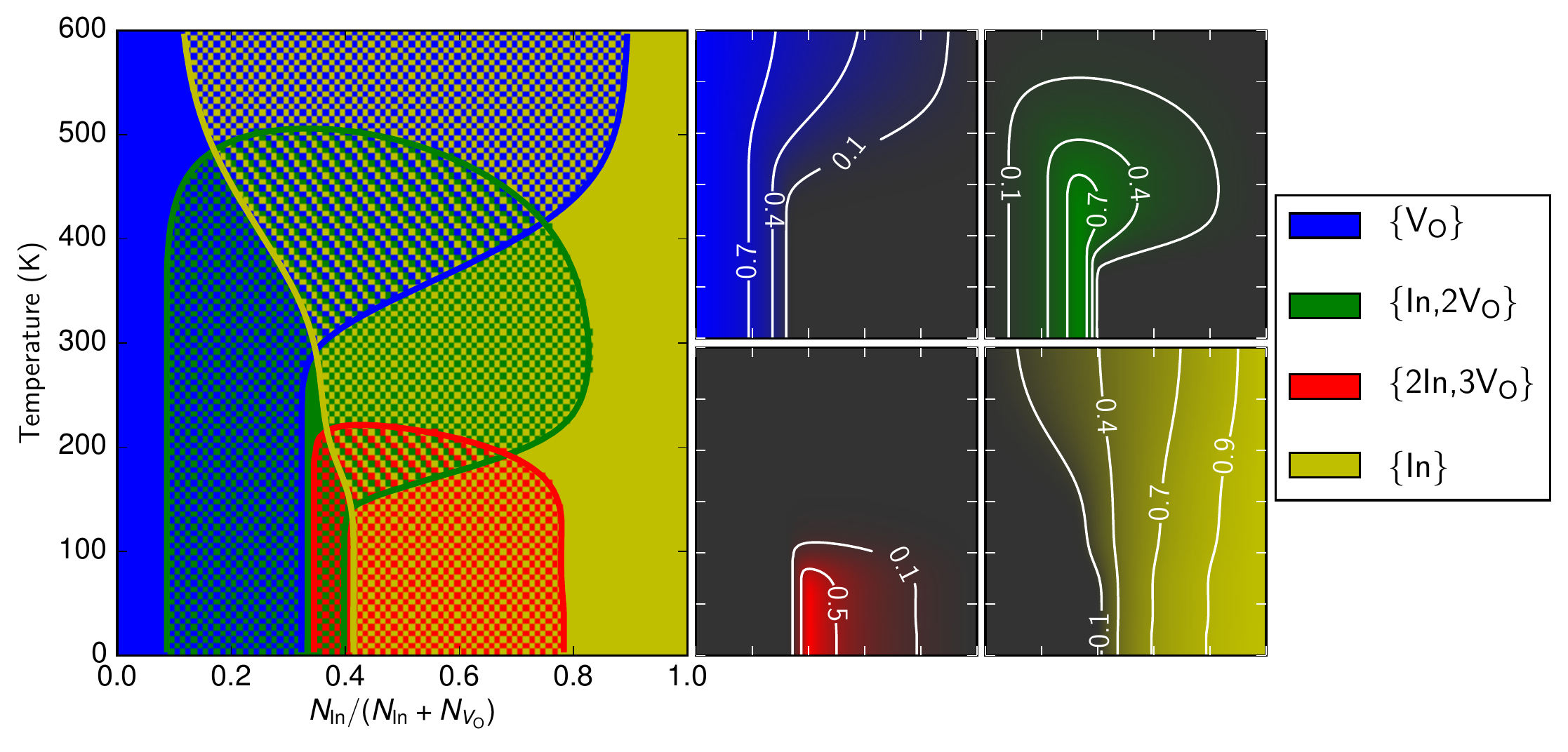}
    \caption{Relative abundances of defect complexes of O vacancies
      and In interstitials in \ino{}.  Contours are plotted on the
      main panel where each defect constitutes 10\% of all defects
      present; likewise, the `checkerboard' pattern indicates which
      defects form at least 10\% of all defects present. On each of
      the smaller panels on the right, the relative abundance of each
      defect is plotted to show more precisely how it varies, along
      with contour lines of equal relative abundance.  In these
      smaller subplots, the colouring, axis ranges and tick marks
      correspond with those on the main panel.}
    \label{fig:oxvacphase}
\end{figure*}

\label{sec:fitemp}
After predicting the defects and their structures, we next predicted their
abundances at finite temperatures using the methods described
previously. Harmonic phonon calculations were performed for all of the
160-atom defect cells and the free energy was constructed and
minimised using Eq.~\ref{eq:futs}. Such calculations were repeated for
a range of temperatures and stoichiometries, giving the defect
abundances as a function of temperature and stoichiometry. This allows
the construction of the phase diagram shown in
Fig.~\ref{fig:oxvacphase} in which the $x$-axis gives the relative
concentration of oxygen vacancies and indium interstitials as in
Fig.~\ref{fig:oxvachull}.

The phase diagram indicates that finite temperature effects are highly
significant, with the \{2In,3\emph{V}$_\mathrm{O}$\} defect
dissociating almost entirely above room temperature. This is because
the configurational entropy is maximised by
breaking composite defects into multiple smaller defects. This
entropically driven dissociation leads, once the
\{2In,3\emph{V}$_\mathrm{O}$\} defects have largely dissociated, to
the formation of a higher number of \{In,2\emph{V}$_\mathrm{O}$\}
defects over certain temperature ranges, until the temperature
increases still further and entropy maximisation dominates over energy
minimisation so that the defect complexes break up completely and the
\{\emph{V}$_\mathrm{O}$\} and \{In\} point defects no longer bind to each
other.  (We have also recomputed the phase diagram using only those
defects which had appeared in the literature prior to this paper, and
the results confirm that our new defects significantly alter the
predicted interaction between oxygen vacancies and indium
interstitials; the recomputed phase diagram is shown in the
Supplemental Material.)

We find the impact of the orientational degeneracy term to be very
significant: by recomputing the phase diagram with translational
entropy included but orientational entropy ignored, we calculate that
the inclusion of the orientational degeneracy term pushes the
temperature above which the \{In,2\emph{V}$_\mathrm{O}$\} defect
constitutes under 10\% of all defects from \SI{820}{\kelvin} down to
the \SI{500}{\kelvin} shown in Fig.~\ref{fig:oxvacphase}. This can be
understood once again as an entropically driven destabilisation of larger defect
complexes such as \{In,2\emph{V}$_\mathrm{O}$\} relative to their
dissociation into smaller defects. 
We note that the magnitude of
this change -- a full \SI{320}{\kelvin} shift in the transition
temperature at which \{In,2\emph{V}$_\mathrm{O}$\} dissociates --
demonstrates the importance of capturing the effects of orientational
entropy, as we have done here using our group theory scheme.
(The recalculated phase diagram without orientational entropy included is also shown in the Supplemental Material.)

The defect structures, formation energies, degeneracies, and sources
(AIRSS or literature) are summarised in
Table~\ref{tab:structures}. Note that the large number of defects with
a degeneracy of 24 arises because the primitive cell of \ino{} has 24
symmetry operators, and most of the defects break all of the
symmetries of the bulk.

In conclusion, we have implemented a general methodology for
first-principles predictions of defect abundances in crystalline
materials at finite temperature. This methodology has been applied to
bixbyite \ino{}. Vibrational corrections and defect degeneracies are
included in the free energies, which has led to the prediction of
novel defect complexes from first principles.  Using a
group theory method, we have calculated the important but
often-neglected contribution of the orientational degeneracy to the
defect free energies, which has allowed us to provide a more accurate
prediction of the abundances of the defects than has been possible
to obtain before.

Studies of the intrinsic doping of \ino{} have focused on the oxygen
vacancy and indium interstitial defects as possible causes, but the
results of our structure searches suggest that these point defects in
fact tend to bind to each other to form composite defects whose
structures we have described here for the first time; moreover, our
finite-temperature calculations suggest that they remain bound at
ambient conditions. By calculating the electronic DOS 
of these defects, we have also shown that they have the potential to
act as shallow donors, which provide a source of charge
carriers. These results could have significant implications for the
question of the cause of intrinsic doping in \ino{}, whose origin has
been the subject of ongoing controversy. Given experimental findings
that the conductivity of \ino{} varies significantly with its
annealing temperature, our calculations could also offer greater
insight into the underlying defect physics affecting the manufacturing
process.

\begin{table}[t]
\begin{tabular}{ M{1.7cm} | M{1.7cm} | M{1.6cm} | M{1.5cm} }
\hline\hline
Defect & $E_F$/atom & Degeneracy & Origin \\ \hhline{-|-|-|-}
\{In,2V$_\mathrm{O}$\} & -0.16 & 24 & AIRSS \\ 
\{In,V$_\mathrm{O}$\} & -0.13 & 24 & Known~\cite{Agoston2009} \\ 
\{In,V$_\mathrm{O}$\}* & \ 0.00 & 24 & AIRSS \\ 
\{2In,2V$_\mathrm{O}$\} & -0.05 & 24 & AIRSS \\ 
\{In\} & \ 0.00 & 4 & Known~\cite{Tomita2005} \\ 
\{2In,3V$_\mathrm{O}$\}* & \ 0.00 & 24 & AIRSS \\ 
\{V$_\mathrm{O}$\} & \ 0.00 & 24 & Known~\cite{Tanaka} \\ 
\{2In,3V$_\mathrm{O}$\} & -0.17 & 24 & AIRSS \\
\hline\hline
\end{tabular}

\caption{Each defect appearing in this study is reported, along with its 
  $T=0$ formation energy per atom, orientational degeneracy, 
  and origin.  All of the structures in this Table were found by AIRSS, 
  but those marked as `AIRSS' are believed to have been found in this 
  study for the first time, whereas those whose origin is `Known' have 
  appeared previously in the literature. The formation energies of 
  \{\emph{V}$_\mathrm{O}$\} and \{In\} are zero by definition, since we 
  have used them as the reservoirs of oxygen vacancies and indium 
  interstitials.}
\label{tab:structures}
\end{table}

\FloatBarrier

\begin{acknowledgments}
  R.\ J.\ N.\ acknowledges financial support from the Engineering and
  Physical Sciences Research Council (EPSRC) of the U.K.\
  [EP/J017639/1].  
  This work was performed using the Darwin Supercomputer of the
  University of Cambridge High Performance Computing Service
  (http://www.hpc.cam.ac.uk/), provided by Dell Inc. using Strategic
  Research Infrastructure Funding from the Higher Education Funding
  Council for England and funding from the Science and Technology
  Facilities Council.
  R.\ J.\ N.\ and A.\ J.\ M.\ acknowledge use of the Archer facilities
  of the U.K.'s national high-performance computing service (for which
  access was obtained via the UKCP consortium EP/K014560/1). J.\ M.\
  W.\ acknowledges the support of the EPSRC Centre for Doctoral
  Training in Computational Methods for Materials Science.
  A.\ J.\ M.\ acknowledges support from the Winton Programme for
  the Physics of Sustainability.

  Data used in this work are available via the Cambridge data
  repository at dx.doi.org/10.17863/CAM.4396.

  We are grateful to Chris G. Van De Walle for fruitful discussions.

\end{acknowledgments}

\bibliography{references.bib}
\end{document}


\title{Structures and finite-temperature abundances of defects in
  In$_2$O$_3$-II from first-principles calculations: supplementary information} \author{Jamie
  M. Wynn} \author{Richard J. Needs} \author{Andrew J. Morris}
\affiliation{Theory of Condensed Matter Group, Cavendish Laboratory,
  University of Cambridge}
\maketitle
\subsection{Structure searching} 
All density-functional theory (DFT) calculations were performed using version 8 of the \textsc{castep}~\cite{Clark2005} plane-wave code. In the initial structure search we used Vanderbilt ultrasoft pseudopotentials~\cite{Vanderbilt1990} generated by \textsc{castep}. In these pseudopotentials the $4d$, $5s$ and $5p$
electrons in indium were treated as valence electrons and the others as core
electrons, while for oxygen the $2s$ and $2p$ electrons were treated
as valence electrons\footnote{The \textsc{castep} pseudopotential generation strings used to generate the pseudopotentials were \texttt{3|2.3|6|8|8|50:51:42}
for indium, and
\texttt{2|1.1|14|16|19|20:21(qc=7)}
for oxygen.}.

\begin{figure}
\centering
\includegraphics[width=0.5\columnwidth]{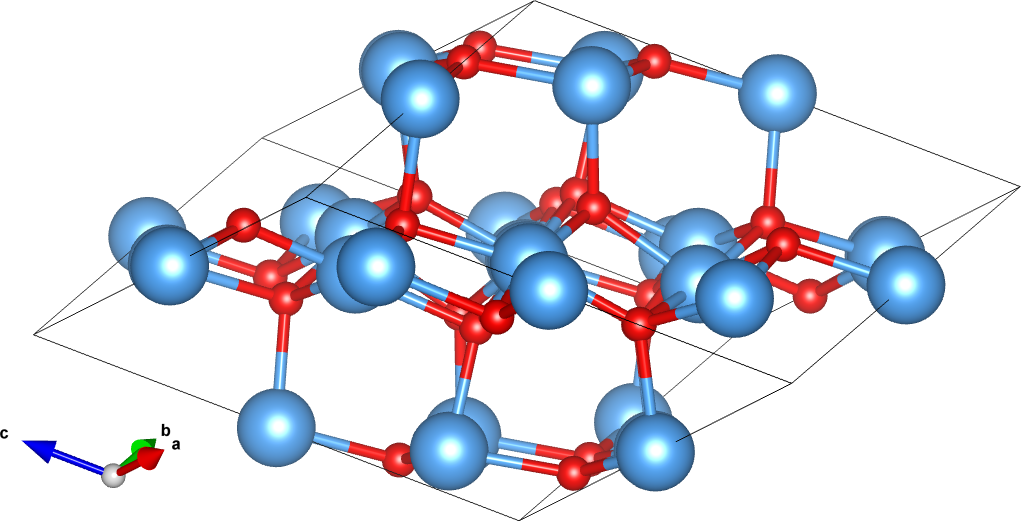}
\caption{The 40-atom primitive unit cell of \ino{} used for structure
  searching in this work. Oxygen atoms are in red, and indium atoms
  are blue. The structure is shown along the [112] direction, which
  reveals it to be slightly displaced from a series of planes of O and
  In atoms.}
\label{fig:in2o3stable}
\end{figure}

By performing a series of total energy calculations, we found that
a cutoff of \SI{500}{\electronvolt} is
sufficient to obtain formation energies to within 
\SI{12}{\milli\electronvolt} which is accurate enough for the purposes of the initial AIRSS
searches.
We performed the searches in the 40-atom primitive unit cell shown in Fig.~\ref{fig:in2o3stable}.
The Brillouin zone (BZ) was sampled using only the Baldereschi mean-value point~\cite{Baldereschi1973}, which we find is associated with a sampling error of less than \SI{0.02}{\electronvolt} per atom. We used the lattice constant obtained from the PBE functional of
\SI{10.21}{\angstrom} which is in agreement with the PBE results
presented in Ref.~\citealp{Varley2015} and is 0.8\% larger than the
experimental value~\cite{Marezio1966}.

\subsection{Supercell calculations}
After the initial structure searching phase, the structures of the defects lying closest
to the convex hull were refined by re-optimising them in a larger cell
with more accurate computational settings. We used a 160-atom
supercell with the shape chosen to maximise the distance from each defect to its nearest
periodic image.  This supercell is the primitive cell of a
face-centred cubic lattice, with lattice parameters
$a=b=c=\SI{14.44}{\angstrom}$ and angles $\alpha=\SI{60}{\degree}$,
$\beta=\SI{60}{\degree}$ and $\gamma=\SI{90}{\degree}$.

The geometry optimisation in the 160-atom supercell was performed with a higher basis set cutoff of \SI{700}{\electronvolt} and BZ sampling on a $2\times2\times2$ Monkhorst-Pack grid. We use the same ultrasoft pseudopotentials as for the initial search.

We calculate (via calculations performed at still higher basis cutoffs) that truncating the plane wave basis beyond \SI{700}{\electronvolt} results in an error in the formation energies of under \SI{2}{\milli\electronvolt} per atom. Some calculations performed with a $4\times 4\times 4$ MP grid indicate that the corresponding error in the total energy due to the BZ sampling is under \SI{4}{\milli\electronvolt} per atom.

To calculate the vibrational free energy of a defect, we first relax the 160-atom supercell containing that defect so that the forces on the atoms are no larger than \SI{0.01}{\electronvolt\per\angstrom}. We then perform harmonic phonon calculations using the finite-difference method, solving the Kohn-Sham equations at $\Gamma$ only (due to the significant cost of
phonon calculations on such large cells and the fact that the phonon
free energy is generally a small correction). The elements of the dynamical matrix
were evaluated using the finite displacement method, and only phonons
with wavevector $\mathbf{q}=0$ were considered. 

\subsection{Metastable defects}
\begin{figure}
    \centering
    \includegraphics[width=0.3\columnwidth]{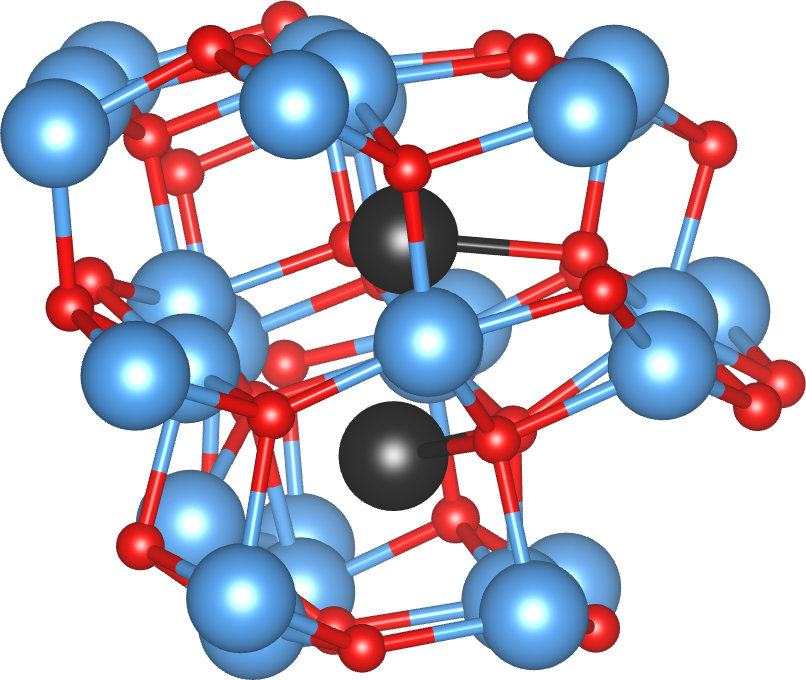}
    \caption{The \{2In,2\emph{V}$_\mathrm{O}$\} defect: two In
    interstitials lie either side of a layer of In atoms, with each
    of them forming ionic bonds with two nearby O atoms.}
    \label{fig:2in2vo}
\end{figure}

\begin{figure}
    \centering
\hfill
    \subfloat[]{
    \includegraphics[width=0.22\columnwidth]{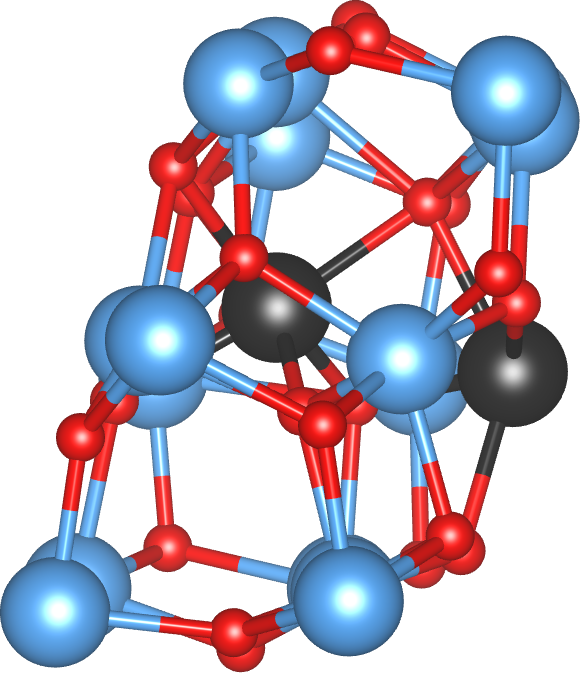}
}
\hfill
    \subfloat[]{
    \includegraphics[width=0.36\columnwidth]{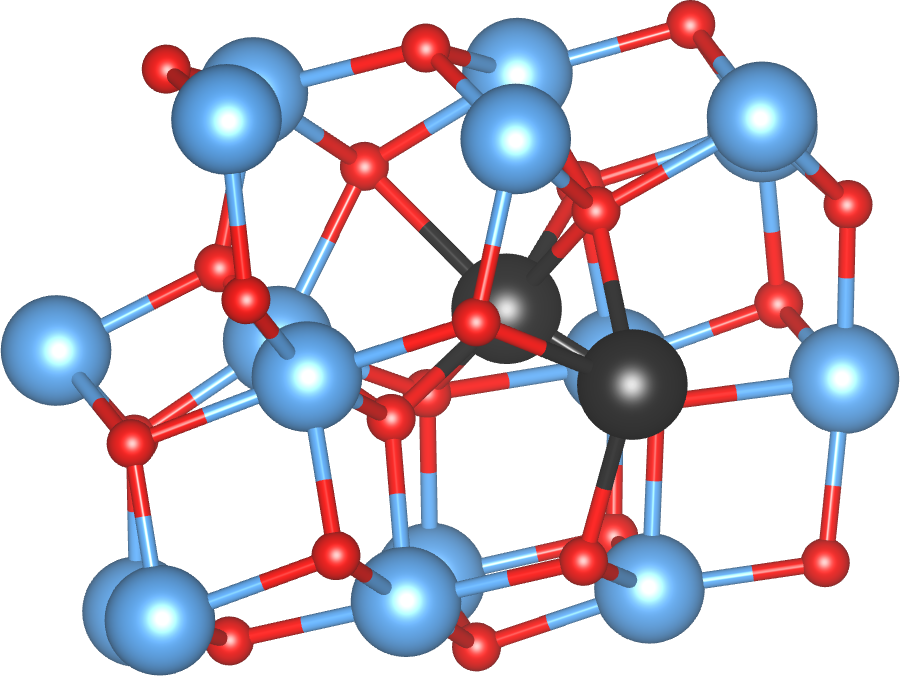}
}
\hfill
\null
    \caption{The \{In,\emph{V}$_\mathrm{O}$\}* defect: an In split interstitial (dark black), combined with significant repositioning of nearby oxygen atoms.}
    \label{fig:invostar}
\end{figure}

\begin{figure}
    \centering
\hfill
    \subfloat[]{
    \includegraphics[width=0.3\columnwidth]{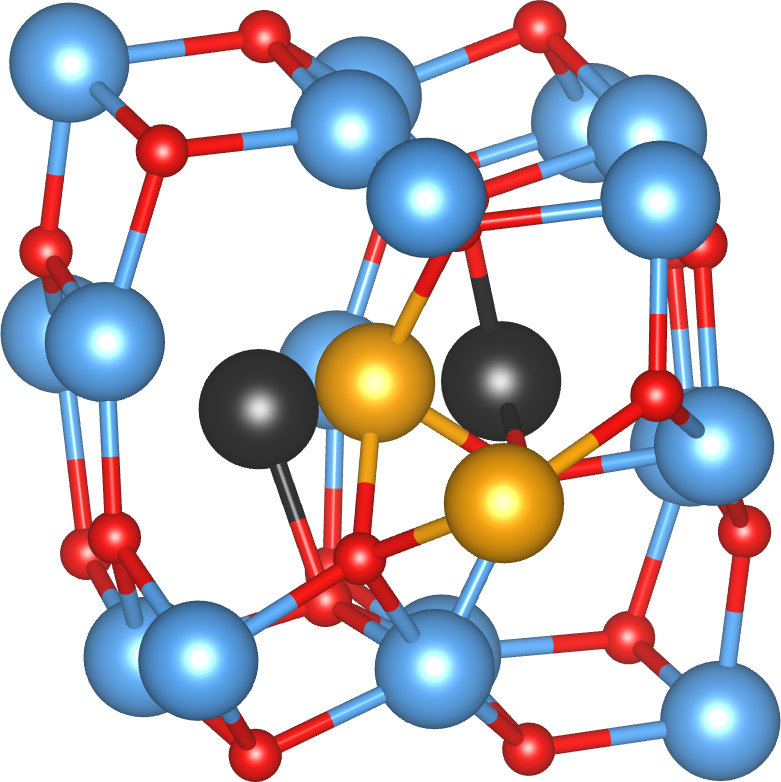}
}
\hfill
    \subfloat[]{
    \includegraphics[width=0.3\columnwidth]{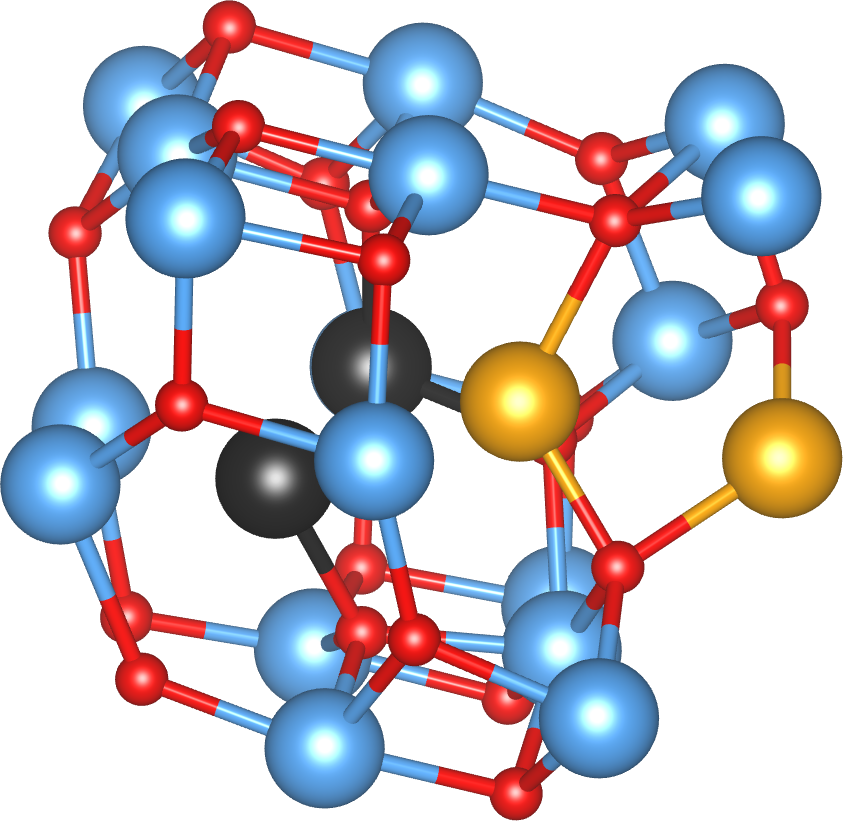}
}
\hfill\null
    \caption{The \{2In,3\emph{V}$_\mathrm{O}$\}* defect: two In split interstitials (one in black, one in orange) are positioned so that the orange interstitial points into the site about which the first interstitial forms.}
    \label{fig:2in3vostar}
\end{figure}

Because their relative energetic unfavourability renders them of lesser physical importance, three metastable defect structures were not shown in the Letter. Nevertheless, they are novel structures and are predicted to be stable relative to dissociation into isolated indium interstitials and oxygen vacancies, so we show them here. In the \{2In,2\emph{V}$_\mathrm{O}$\} (pictured in
Fig.~\ref{fig:2in2vo}), two In interstitials (black) sit above and below a
layer of In atoms, and each binds to two nearby O atoms (similarly
to \{In,2\emph{V}$_\mathrm{O}$\}). The \{In,\emph{V}$_\mathrm{O}$\}* defect is a split indium interstitial shown on Fig.~\ref{fig:invostar}, whilst the larger \{2In,3\emph{V}$_\mathrm{O}$\}* complex consists of two split In interstitials in close proximity, and is shown on Fig.~\ref{fig:2in3vostar}.

\subsection{Finite-temperature abundances}
\begin{figure}
    \raggedright
    \hspace{7.4em}
    \includegraphics[width=0.65\textwidth]{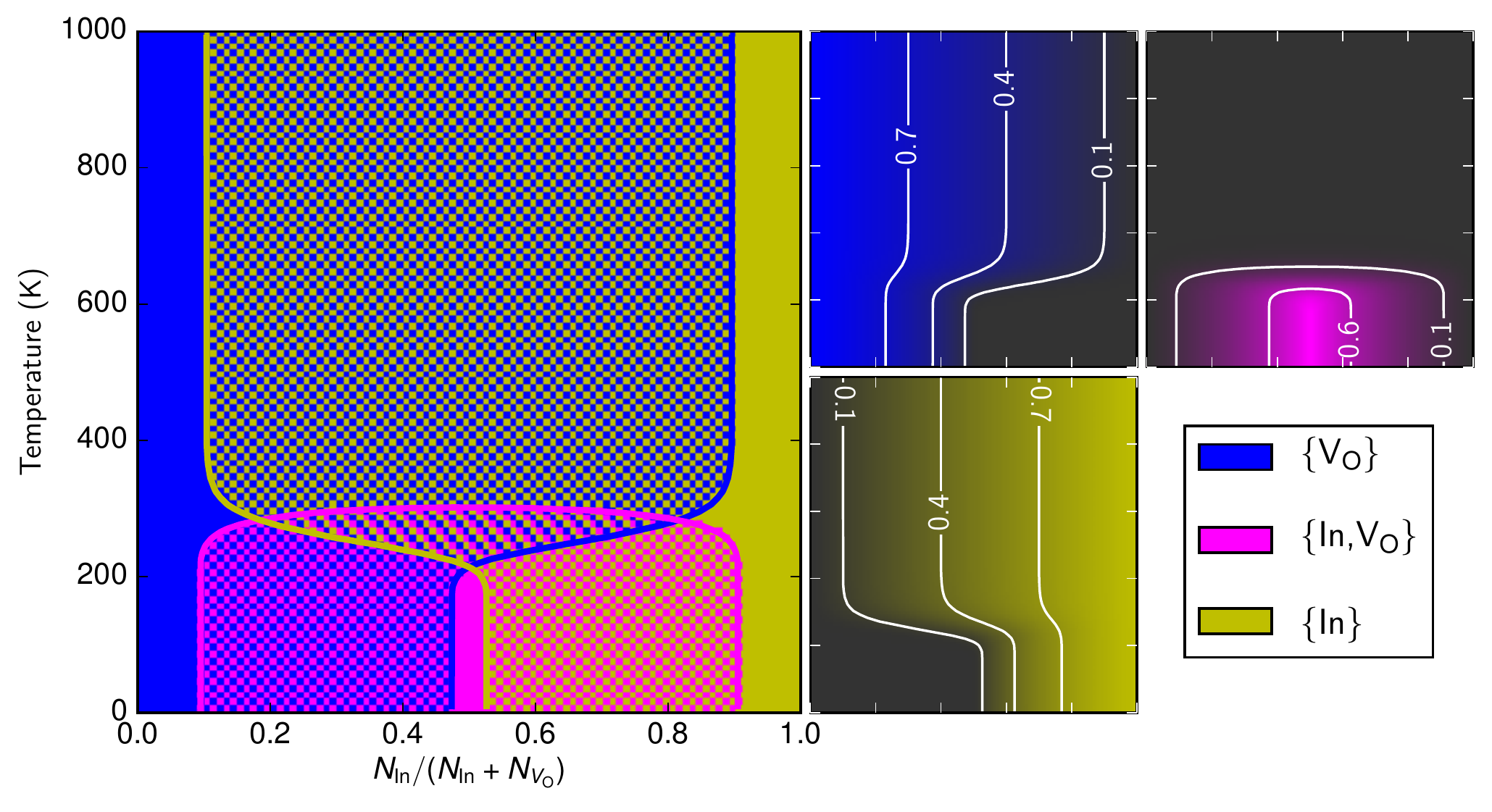}\\
    \hspace{7em}
    \includegraphics[width=0.80\textwidth]{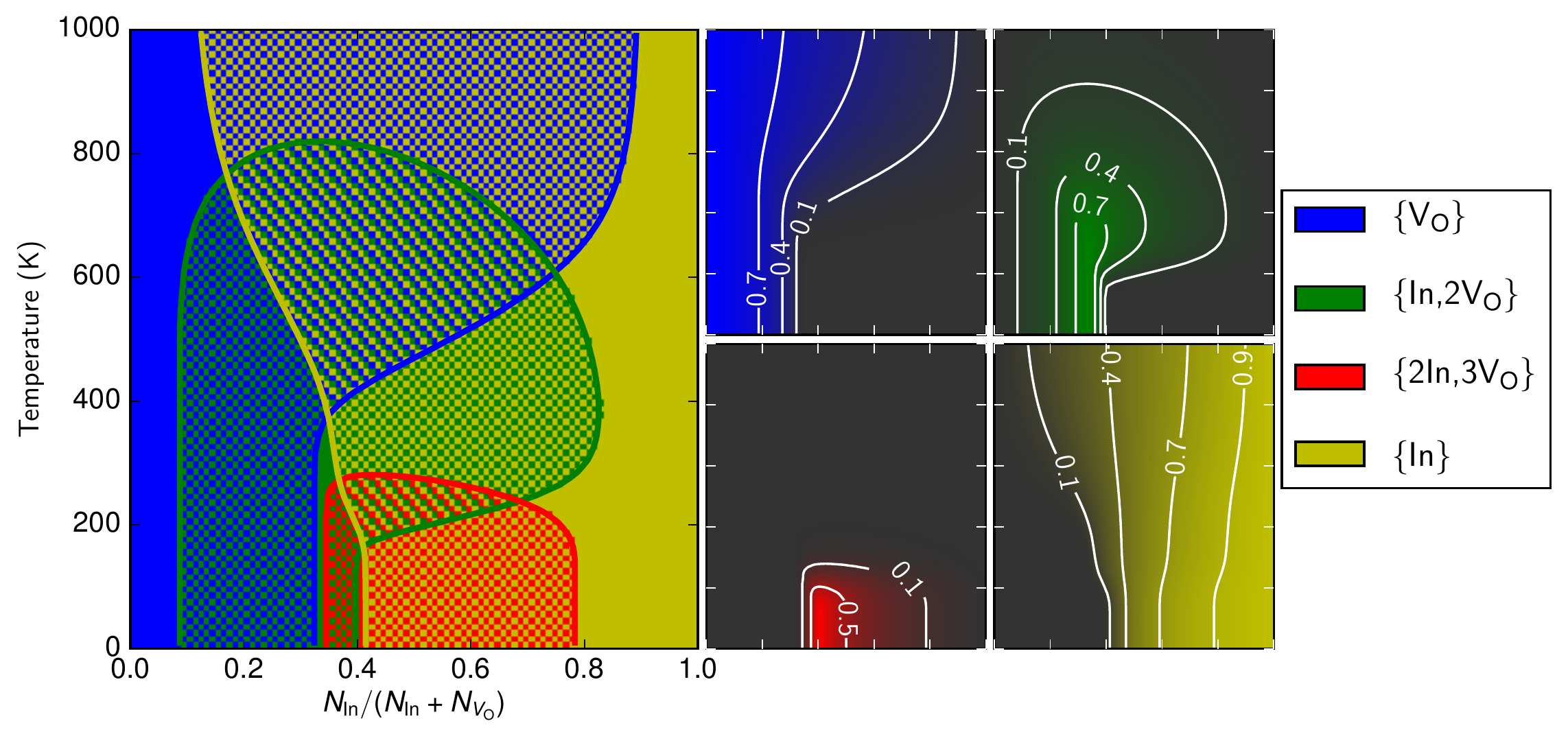}\\
    \hspace{6.85em}
    \includegraphics[width=0.81\textwidth]{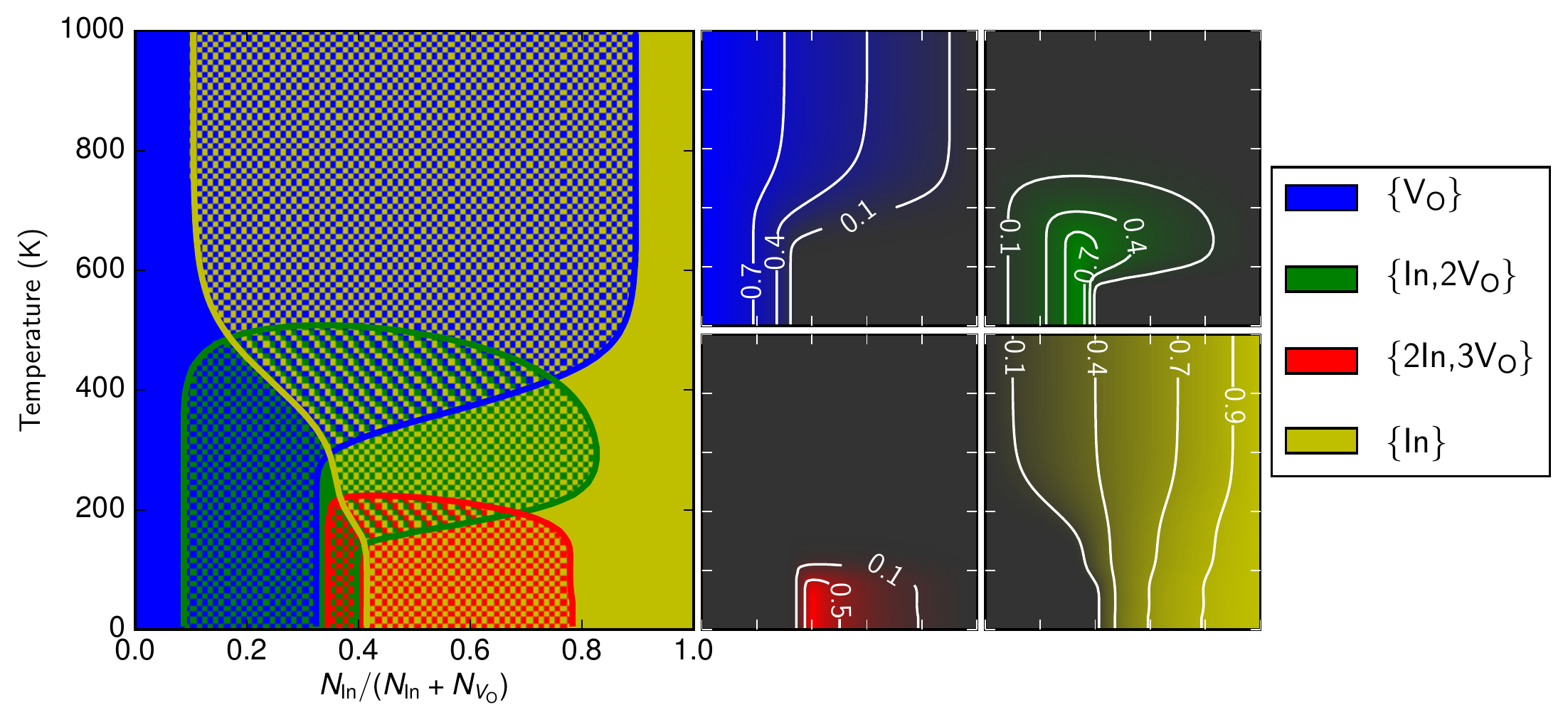}
    \caption{The defect phase diagram calculated using only previously known defects (top), including newly found defects but ignoring the effect of orientational entropy (middle), and -- for comparison -- including both new defects and orientational entropy (bottom), as in Fig. 4 of the main text. As with the phase diagram in the main text, the smaller subpanels show the relative abundance of each defect along with contours of equal relative abundance.
    }
    \label{fig:oxvacphase}
\end{figure}

To illustrate the effects of both the novel defects we have discovered, and of our inclusion of the configurational entropy term, we have recalculated the phase diagram without each of these factors. The resulting phase diagrams (Fig.~\ref{fig:oxvacphase}) show markedly different behaviour: the simple \{In,\emph{V}$_\mathrm{O}$\} indium substitutional is not as energetically favourable as the new \{2In,3\emph{V}$_\mathrm{O}$\} and \{In,2\emph{V}$_\mathrm{O}$\} defects, and as a result it has largely dissociated at room temperature. The significance of the configurational entropy term is also clear on Fig.~\ref{fig:oxvacphase}, with a marked decrease in the stability of the composite \{2In,3\emph{V}$_\mathrm{O}$\} and \{In,2\emph{V}$_\mathrm{O}$\} defects when the effects of configurational entropy are taken into account.

\subsection{HSE06 calculations}

\begin{figure}
    \centering
    \includegraphics[width=0.5\columnwidth]{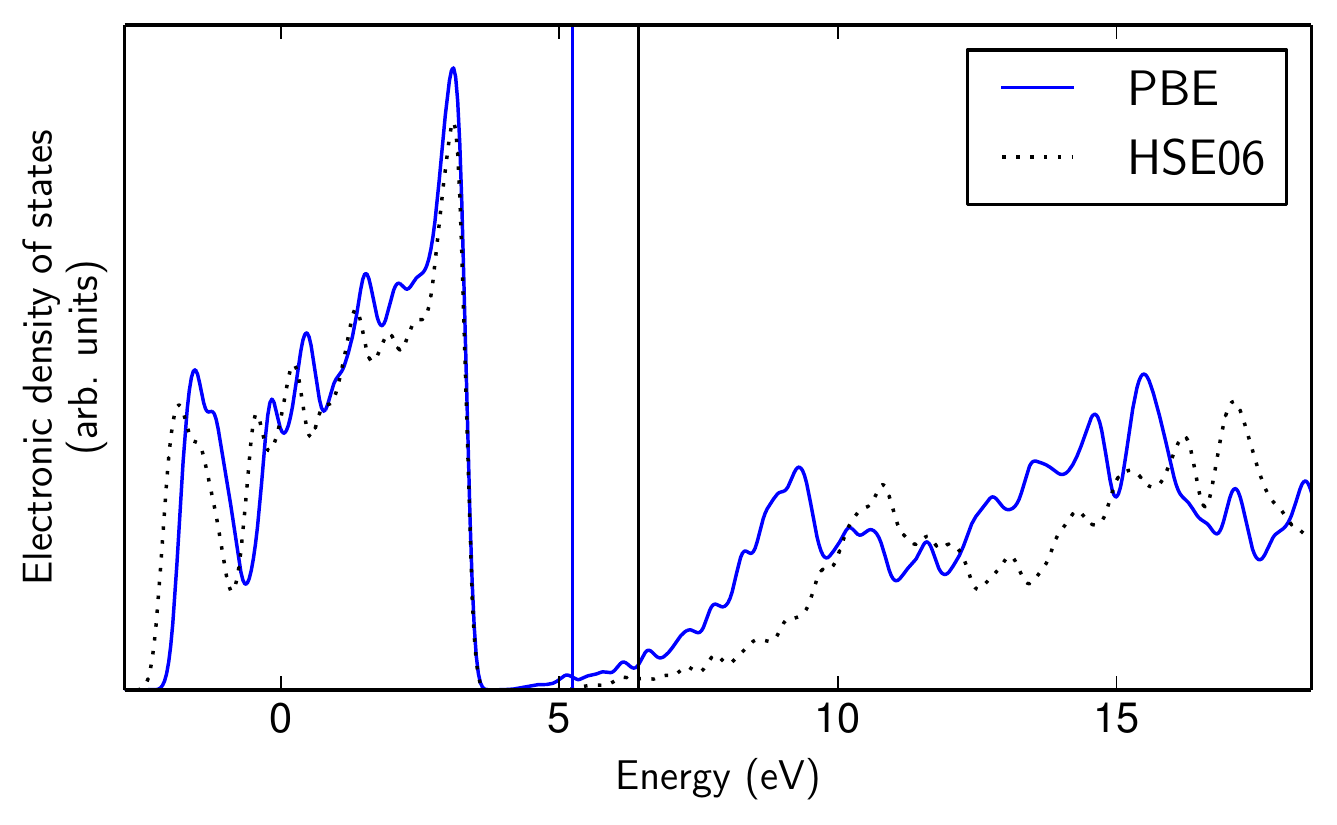}
    \caption{Electronic density of states in the vicinity of the band
      gap for the \{$\emph{V}_\mathrm{O}$\} defect, computed with the
      PBE and HSE06 functionals. The densities of states have been
      shifted to align their VBMs, and the Fermi levels for the two
  functionals are indicated by vertical lines.}
        \label{fig:hse}
\label{fig:oxvacdos}
\end{figure}

In our HSE06 density of states calculations we use the conventional cubic 80-atom unit cell with the Kohn-Sham equations solved
self-consistently at the $\Gamma$ point of the BZ only (because of the extreme cost of computing the screened exchange term), before solving
the Kohn-Sham equations non-self-consistently on a $5\times 5\times 5$
Monkhorst-Pack grid. (We choose an odd $n$ for the MP grid because \ino{}-II has a direct band gap at $\Gamma$; this would be a suboptimal choice for the calculation of total energies, but for densities of states it gives more accurate energies for the valence band maximum and conduction band minimum.) Norm-conserving pseudopotentials were used which treated as valence the indium $4d$, $5s$ and $5p$ orbitals and the oxygen $2s$ and $2p$ orbitals. We increased the plane wave
basis set cutoff to \SI{850}{\electronvolt} for these calculations. The resulting density of states is shown alongside that resulting from the PBE functional is shown on Fig.~\ref{fig:hse}; both densities of states are calculated using version 1.2 of the OptaDOS \cite{Yates2007,Morris2014} code.

\FloatBarrier

\bibliography{references.bib}